\pdfoutput=1

\documentclass[11pt]{article}

\usepackage{acl}

\usepackage{times}
\usepackage{latexsym}
\usepackage{todonotes}
\usepackage{multirow}
\usepackage{makecell}
\usepackage{booktabs}
\usepackage{tablefootnote}
\usepackage{mathtools, nccmath}
\usepackage{amsmath}
\usepackage{graphicx,subcaption}
\usepackage{float}
\usepackage{soul}
\usepackage{colortbl} 
\NewDocumentCommand\emojiduck{}{
    \includegraphics[scale=0.07]{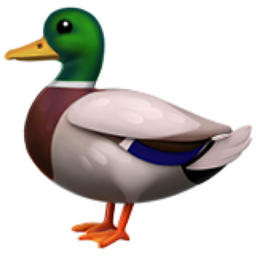}}
\newcommand{\relup}[1]{\textcolor{blue}{(+#1\%)}}
\newcommand{\reldown}[1]{\textcolor{red}{(-#1\%)}}

\DeclarePairedDelimiter\floor{\lfloor}{\rfloor}
\DeclareMathOperator*{\argmax}{argmax}
\DeclareMathOperator*{\argsort}{argsort}

\usepackage[T1]{fontenc}

\usepackage[utf8]{inputenc}

\usepackage{microtype}

\usepackage{inconsolata}
\usepackage{float}

\newcommand{\OurMethod}{\textbf{DUQGen}}
\newcommand{\OurMethodRerank}{\textbf{DUQGen-reranker}}
\newcommand{\OurMethodFirstStage}{\textbf{DUQGen-retriever}}

\title{\emojiduck~\OurMethod{}: Effective Unsupervised Domain Adaptation of Neural Rankers by Diversifying Synthetic Query Generation}

\author{Ramraj Chandradevan, Kaustubh D. Dhole,  Eugene Agichtein \\
        Department of Computer Science \\ Emory University \\ Atlanta, USA-30307\\
        \texttt{\{rchan31,kdhole,yagicht\}@emory.edu}}

\begin{document}
\maketitle
\begin{abstract}

State-of-the-art neural rankers pre-trained on large task-specific training data such as MS-MARCO, have been shown to exhibit strong performance on various ranking tasks without domain adaptation, also called zero-shot. However, zero-shot neural ranking may be sub-optimal, as it does not take advantage of the target domain information. Unfortunately, acquiring sufficiently large and high quality target training data to improve a modern neural ranker can be costly and time-consuming. To address this problem, we propose a new approach to unsupervised domain adaptation for ranking,~\OurMethod, which addresses a critical gap in prior literature, namely how to automatically generate both effective and diverse synthetic training data to fine tune a modern neural ranker for a new domain. Specifically,~\OurMethod{} produces a more effective representation of the target domain by identifying clusters of similar documents; and generates a more diverse training dataset by probabilistic sampling over the resulting document clusters. Our extensive experiments, over the standard BEIR collection, demonstrate that~\OurMethod{} consistently outperforms all zero-shot baselines and substantially outperforms the SOTA baselines on 16 out of 18 datasets, for an average of 4\% relative improvement across all datasets. We complement our results with a thorough analysis for more in-depth understanding of the proposed method's performance and to identify promising areas for further improvements.

\end{abstract}

\section{Introduction}

Large Language Models (LLMs) have enabled new state-of-the-art performance in neural ranking~\cite{yan2019idst, kamps2020impact,p3ranker,nogueira2020document}. An effective approach has been to train the LLMs on a large-scale general ranking task such as MS-MARCO passage or document ranking~\cite{ms-marco} or Wikipedia retrieval~\cite{clirmatrix}, to learn task-specific features, which are often shared across other domains and datasets. The resulting rankers can then be used, without any adaptation (or in a zero-shot way) for a wide range of ranking tasks. For example, the BEIR~\cite{beir} benchmark had demonstrated SOTA or near-SOTA performance of several zero-shot neural rankers on a diverse set of retrieval tasks.

\begin{figure}[t!]
\begin{center}
    \includegraphics[width=\columnwidth]{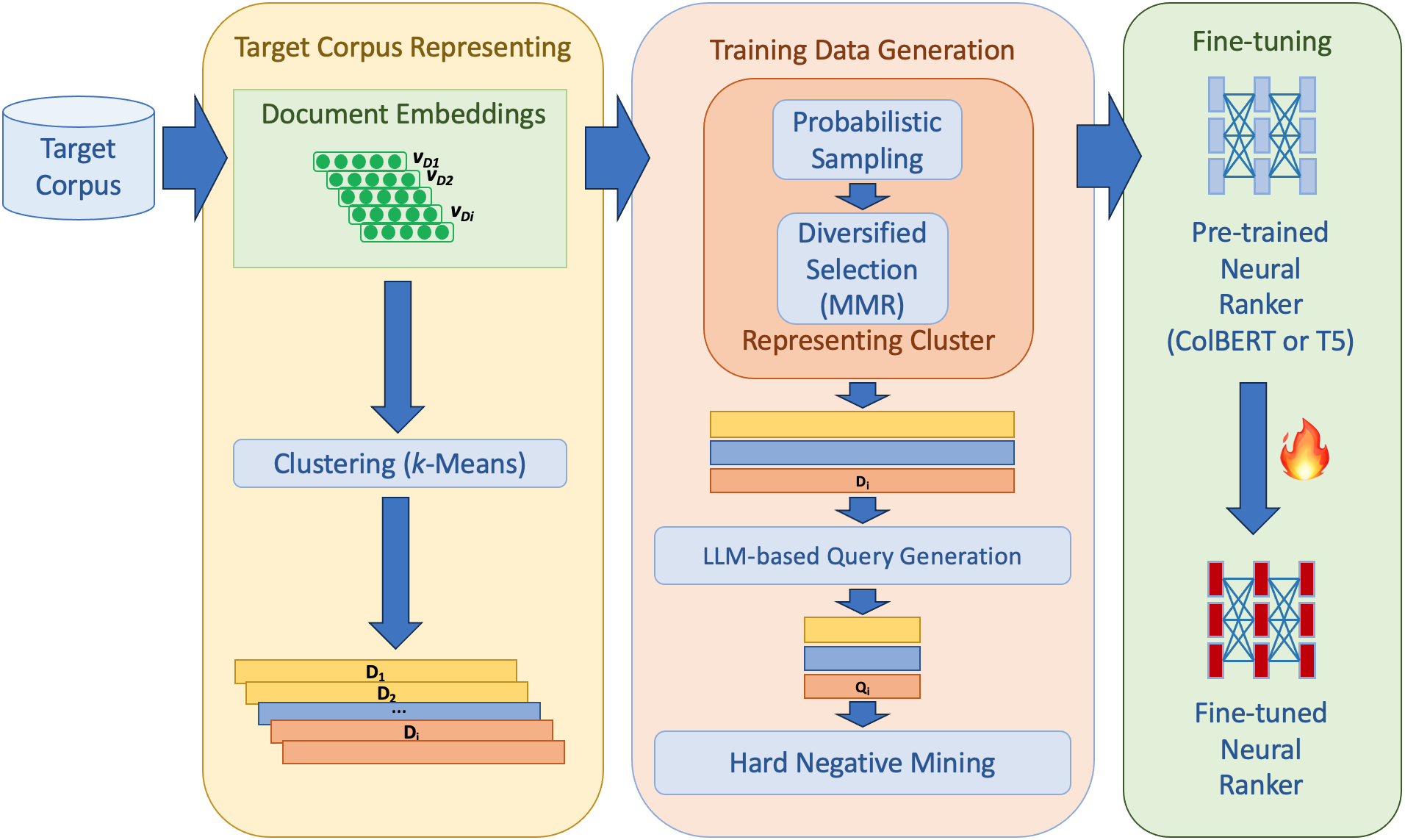}
    \caption{\OurMethod: an unsupervised domain-adaptation framework for neural ranking.}
    \label{fig:pipeline}
    \vspace{-15pt}
\end{center}
\end{figure}

However, when switching to specialized domains such as finance or scientific documents, zero-shot ranking performance should benefit from additional information for the target domain. For training modern neural rankers, acquiring sufficiently large and high-quality target training data of query-document pairs, to improve a modern neural ranker, can be costly and time-consuming. Hence, there has been significant interest in various approaches to domain adaptation for neural rankers, with varying degrees of supervision, including unsupervised approaches using synthetically generated queries, documents, or both query-document pairs~\cite{UPR, inpars,inparsv2,promptagator,docgen}.

Unfortunately, most of the previously reported results did not exceed the ranking performance compared to the current SOTA zero-shot models, as evaluated on the BEIR benchmark~\cite{beir}. In other words, to the best of our knowledge, {\em no previously reported unsupervised ranking adaptation method demonstrated consistent improvements over large neural SOTA zero-shot rankers.}

In this work, we investigate whether it possible to improve the ranking performance of a pre-trained SOTA neural ranker for a given target domain, through unsupervised domain adaptation (UDA). To address this question, we first identify a critical requirement for the synthetic training data to be effective for ranking adaptation: that the generated training data should be both {\em representative of the target domain}, and {\em sufficiently diverse} to force changes to the ranking model at the appropriate representation level, without causing over-fitting or catastrophic forgetting (i.e., degrading performance on the original ranking tasks).

Specifically, we propose a new method~\OurMethod, which stands for \textbf{D}iversified \textbf{U}nsupervised \textbf{Q}uery \textbf{Gen}eration. ~\OurMethod introduces a general approach for ranking domain adaptation, which focuses on selecting representative and diverse set of documents and query pairs for training a neural ranker.~\OurMethod{} only requires access to a (part of) target document collection to be searched, and can improve any pre-trained neural ranker.~\OurMethod{} is illustrated in Figure~\ref{fig:pipeline}, and introduces the following innovations compared to previous unsupervised ranking adaptation approaches: (1) representing the target document collection as document {\em clusters};  (2) diversifying the synthetic query generation by probabilistic sampling over the resulting document clusters; and (3) prompting a large LLM for query generation with in-context examples to generate queries from the selected documents. As we show experimentally, these innovations are responsible for consistent improvements over the previous SOTA baselines for ranking adaptation on almost all BEIR benchmarks, as well as consistent improvements over zero-shot performance of SOTA neural rankers. In summary, our contributions include:

\begin{enumerate}
    \item \OurMethod, a general and effective unsupervised approach for domain adaption of neural rankers via synthetic query generation for training.
    \item A novel and general method for creating representative and diverse synthetic query data for a given collection via clustering and probabilistic sampling.
    \item Comprehensive experiments demonstrating that~\OurMethod{} consistently outperforms all SOTA baselines on 16 out of 18 BEIR datasets, and thorough analysis of the components of~\OurMethod{} responsible for the improvements. We release all our code and models publicly\footnote{\href{https://github.com/emory-irlab/DUQGen }{https://github.com/emory-irlab/DUQGen}\label{duqgen_code}}.
\end{enumerate}

Next, we describe prior work on domain adaptation of neural rankers in more detail, to place our contributions in context.

\section{Related Work}
\label{related_work}

In this section, we discuss the prior works that help to establish the problem and navigate the solution.

\subsection{Neural Rankers}

Recently, transformer-based pre-trained language models have demonstrated impressive effectiveness in neural rankers~\cite{bert-and-beyond}. A neural ranker returns an order list of documents given a query, where a relevancy score between query and document dense embedding representations is used for sorting. Extensive studies have been conducted on both dense retrievers and re-rankers~\cite{neuralIR-survey}. In comparison to encoder-based rankers (MonoBERT and DuoBERT)~\cite{mono-duo-bert}, encoder-decoder~\cite{monoT5} and decoder-based~\cite{rankllama} rankers exhibit notably superior performance with a larger margin. While ColBERT~\cite{colbert}, Contriever~\cite{contriever}, and GTR~\cite{GTR} perform competitively as dense retrievers, MonoT5-3B~\cite{monoT5} is widely adapted for re-ranking purposes.

\subsection{Unsupervised Domain Adaptation for Neural Rankers}

Despite the remarkable ranking performances demonstrated by recent pre-trained language models in zero-shot settings, they often encounter catastrophic failures in real-world deployment scenarios. The main factor contributing to these failures is \textit{Domain-Shift}~\cite{data-selection} or \textit{Domain Divergence}~\cite{domain-divergences-survey}. Domain-Shift has been a subject of exploration for decades, including recent investigations in domain adaptation~\cite{MS-Shift}. Traditionally, it is assumed that the source and target domains share samples drawn from the same distribution. Previous studies have addressed this issue by quantifying domain divergence through various measures, such as geometric measures, information-theoretic measures, and higher-order measures~\cite{domain-divergences-survey}.
Ultimately, these measures contribute to the development of novel solutions for domain adaptation in neural rankers.

Solutions addressing domain divergence typically fall into two categories: (1) representation learning; and (2) data selection. Representation learning approaches primarily address UDA, with a focus on learning domain-invariant representations~\cite{domain-separation-networks, bhaskar-DA-regularizer-adversarial} or pre-training a zero-shot ranker. On the other hand, data selection assumes that not all samples contribute equally to domain representation~\cite{data-selection-first-work}, highlighting the importance of identifying effective target domain samples. Improper selection of target data during fine-tuning has the potential to undermine the impact of source pre-training. Our research centers on the issue of improper representation of the target domain leading to diminished performances in neural rankers. Therefore,~\OurMethod{} aims to identify representative and diverse target samples that can be effective during the fine-tuning process.

\subsection{Synthetic IR Data Generation}

The increasing power of Large Language models have prompted numerous studies to focus on utilizing LLMs for the creation of high-quality training data. Several previous works have explored unsupervised synthetic data generation for fine-tuning ranking models, including GPL~\cite{GPL}, InPars~\cite{inpars}, InPars-v2~\cite{inparsv2}, DocGen-RL~\cite{docgen}, GenQ~\cite{beir}, and Promptagator~\cite{promptagator}. These frameworks use random document sampling or random seed queries to start their pipelines, which leaves room for improvement.

Each of the previously mentioned works utilizes distinct strategies employed alongside their data synthesis processes. GPL\cite{GPL}, for instance, combines a T5-based~\cite{T5} query generator with a pseudo-labeling cross-encoder to enhance robust learning. InPars and InPars-v2 methods utilize GPT-3 and GPT-J query generators along with different filtering strategies to eliminate low-quality synthetic queries. DocGen-RL introduces an RL-driven guided approach combined with document synthesis using BLOOM~\cite{BLOOM}. GenQ, on the other hand, fine-tunes TAS-B~\cite{TAS-B} with queries generated from an MS-MARCO fine-tuned T5-base generator. 
Promptagator employs a pipeline similar to InPars, but with improved components, such as a random million document samples, a 137B FLAN query generator, and a strong consistency filter to prune 8 million synthetic queries through a relatively complicated and expensive process.
Notably, none of the methods mentioned above take into consideration the significance of identifying domain-representative documents or diversifying the resulting queries. Consequently, the fine-tuned performances appears to fall short of zero-shot performances in many cases.

The quality of the generated training queries significantly affects the end retrieval performances. Despite the utilization of strong query generators (BLOOM and GPT-3), the domain query representation can still be improved.
For instance, InPars employed a prompt containing in-context examples from MS-MARCO training data, yet it still maintains a domain representation gap during in-context generation. Furthermore, their query generation did not address the need for diversity among the generated training samples. Additionally, they incorporated a complex filtering step to prune the generated queries, which we show can be avoided. These methods fine-tuned rankers using large-scale synthetic data, ranging from 100k to 1M examples. In contrast, we argue that judicious selection of training samples can obviate the necessity for such large-scale generation, reducing the required amount of synthetic training data by a factor of x1000.

\section{Methodology}
\label{methodology}


\OurMethod{}, shown in Figure \ref{fig:pipeline}, consists of four components -- domain document selection, domain query generation, negative pairs mining, and fine-tuning. We now cover each component in detail.

\subsection{Domain Document Selection}

We propose to represent a target domain with clusters and each clusters with its sampled documents. Therefore, in this section we describe them in three stages, namely collection document clustering, probabilistic document sampling, and diversified document selection.

\subsubsection{Collection Document Clustering}

Representing a large-scale target collection of documents with limited training data is challenging.
Therefore, we propose to divide the collection into portions, and then sample documents within each portion.
We use a clustering approach for the collection representation.
Moreover, we can achieve diverse topical documents to represent the domain.
We start with the full collection of documents and apply a preprocessing step, where we discard short span documents, filtering out noisy documents.
Then we use a SOTA text encoder, viz. \textit{Contriever} \cite{contriever}, to encode each of the documents.
Using the document embeddings $v_{D_i}$, we apply clustering (\textit{e.g.,} $K$-Means) technique, where $K$ is a hyper-parameter to tune.

\subsubsection{Probabilistic Document Sampling}
\label{sec:prob_sampling}

Representing each cluster within large data collections is challenging since the resultant clusters can often be of imbalanced sizes. Let's take $k^{th}$ cluster size as $c_k$ and collection size as $C$, where ($1 \leq c_k \leq C$).
We ideally want to sample more number of documents from larger size clusters in proportion to the cluster size. If $cluster_k$ and $D_i$ represent $k^{th}$ cluster and its $i^{th}$ document, the probability of selecting $D_i$ from $cluster_k$ is $Pr(D_i|cluster_k) \propto c_k \forall D_i \in cluster_k$.

We intend to sample $N$ number of synthetic training examples from $K$ number of clusters, where $N \geq K$.
Therefore, we design a stratified expression to determine the document sample size $N_k$ for $k^{th}$ cluster, given by

\[N^0_k = 1 + \floor*{\frac{c_k}{C} (N - K)} \]

\[P = N - \sum_{k=1}^{K} N^0_k \]

\[
  N_k =
  \begin{cases}
        N^0_k + 1        & \text{if $k \in \argsort_{top-P} (c_k)$} \\
        N^0_k            & \text{if $k \not\in \argsort_{top-P} (c_k)$}
  \end{cases}
\]

where $N^0_k$ and $P$ are intermediate sample size and integer number. $\floor*{*}$ operation finds the floor integer value.

Now that we determined the sample size for each clusters, we define our sampling approach.
Let's take $d_i$ as the similarity (\textit{e.g. cosine similarity}) between document $D_i$ and its corresponding cluster centroid.
We define an exponential value $e^{d_i}$ as the representative of how close $D_i$ is to its cluster centroid.
Therefore, $Pr(D_i|cluster_k)$ becomes the normalized softmax given by:
\begin{equation}
\label{eq:prob_sampling}
Pr(D_i|cluster_k) = \frac{e^{d_i/T}}{\sum_{j=1}^{c_k} e^{d_j/T} }
\end{equation}
\begin{equation}
\label{eq:cosine_distance}
d_i = cosine(v_{D_i}, \frac{1}{c_k} \sum_{j=1}^{c_k} v_{D_j})
\vspace{-0.2cm}
\end{equation}
where $T$ is the softmax temperature and $v_{D_i}$ is the $i^{th}$ document embedding. Intuitively, a document likelihood to be selected to generate an associated query is proportional to the document similarity to its cluster centroid.

\subsubsection{Diversified Document Selection}
\label{sec: diversified_selection}

Now we sample $N_k$ number of documents from each cluster $cluster_k$ and pool them to obtain the required training size documents $N$. Different sample sets can be drawn from the aforementioned sampling approach with different choices of random seed values.
Therefore, to improve selection robustness in the sampling process, we apply a diversity measure, namely Maximal Marginal Relevance (MMR)~\cite{MMR}.
We first iterate the sampling process $m$ times ($m=5$) to obtain different sample sets.
Then we apply MMR on the pooled documents from $m$ sets to select top-$N_k$ documents for $cluster_k$ as shown:

\begin{equation}
\begin{split}
\argmax_{D_i \in R\setminus S} \Bigl[ \lambda Sim_1 (D_i, D_k) \\
- (1-\lambda) \max_{D_j \in S} Sim_2(D_i, D_j) \Bigr]
\end{split}
\end{equation}

where $D_k$ is the document closest to the cluster centroid, $\lambda$ is a trade-off weight (to be tuned) between similarity to cluster centroid and diversity, $R$ is the pooled documents, $S$ is a subset of documents already selected from $R$, and $Sim_1$ and $Sim_2$ that can be same or different, but we used the $cosine$ similarity for both instances.

\subsection{Synthetic Query Generation}
\label{sec:syn_query_gen}

\begin{figure}[ht]
\begin{center}
    \includegraphics[width=.9\columnwidth]{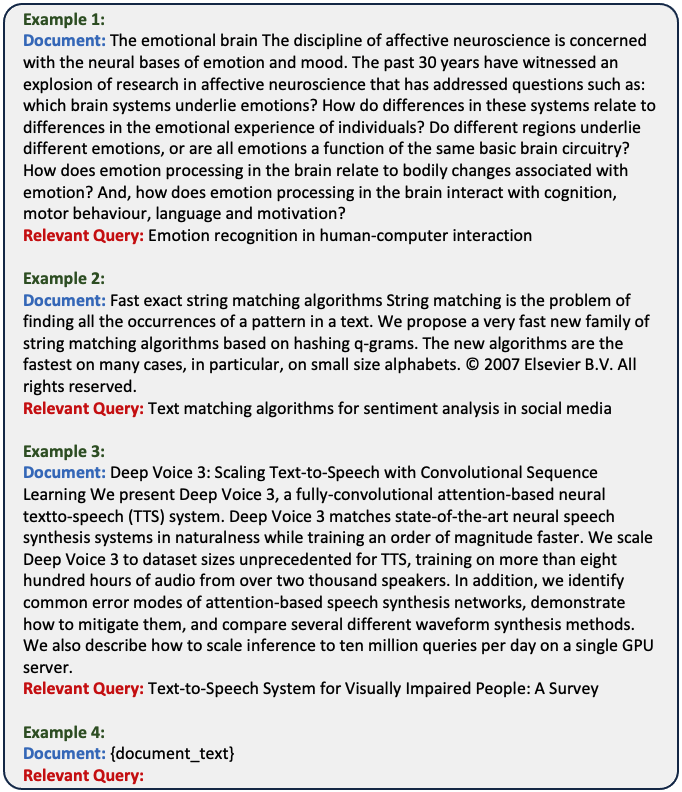}
    \caption{Prompt template with in-context examples for synthetic query generation for the SCIDOCS dataset.}
    \label{fig:scidocs_prompt}
\end{center}
\vspace{-20pt}
\end{figure}

Query generation is an essential component in an unsupervised data generation pipeline for ranking models. Queries represent a target domain \textit{w.r.t.} the user's information need and the domain-task by taking different types, such as questions, headlines, keywords, or claims. Therefore, we use a LLM to generate a synthetic in-domain query for each sampled document. We few-shot prompt the LLM to generate such training queries similar to the existing work of \citet{inpars}. However, our contribution lies in showing that the in-domain few-shot examples (query-document pairs) help to achieve high-quality of queries compared to out-of-domain generic MS-MARCO examples. On each domain, we create a handful (e.g., 3) human generated queries for the few-shot example documents with minimal human effort, and an example prompt is shown in Figure \ref{fig:scidocs_prompt}.

\subsection{Negative Pairs Mining}

After obtaining the domain specific documents and queries, we should generate both positive and negative query-document pairs.
First, the positive query-document pairs can be easily generated by mapping the synthetic queries with their corresponding original (seed) documents.
Second, the negative query-document pairs can be generated from hard negative mining, described in the standard practices~\cite{contriever, ANCE, DPR}.
We parse the synthetic queries to any first-stage retrievers, such as BM25~\cite{bm25}, ColBERT~\cite{colbert}, or Contriever~\cite{contriever}, to get top-$x$ documents.
Then we pick the bottom-$num_{neg}$ documents from the top-$x$ to map against the synthetic queries, where 1:$num_{neg}$ is the \textit{positive:negative} document pair ratio.

\subsection{Fine-tuning with our Synthetic Data}

Our domain adaptation framework can be applied on any ranking models with any weights initialization.
To establish a strong competitor, we leverage the task pre-trained model (on MS-MARCO), and sequentially fully fine-tune with our own generated synthetic data.
We also adapt the same hyper-parameter settings used in the MS-MARCO pre-training stage for fair deliverable.

\section{Experiments}
\label{experiments}
In this section, we provide details of our experimental setup to demonstrate the effectiveness of~\OurMethod{}.

\subsection{Datasets and Metrics}

We employed all 18 datasets from BEIR collection, ranging on diverse retrieval tasks, to assess the effectiveness of our domain adaptation framework on standard out-of-distribution datasets.
Utilizing the \textit{multi-field} index from Pyserini~\cite{Pyserini} for all datasets, we retrieved the top-100 and top-200 documents from lexical and dense first-stage retrievers respectively.
Subsequently, we restricted re-ranking to top-100 BM25 documents and top-200 dense retriever documents. Since we evaluate our approach on both first-stage retrieval and re-ranking, we measured both nDCG@10 and R@100.

\subsection{Ranking Models}

We fine-tuned ColBERT\footnotemark{}\footnotetext{https://github.com/stanford-futuredata/ColBERT}~\cite{colbert} and MonoT5-3B\footnotemark{}\footnotetext{castorini/monot5-3b-msmarco-10k}~\cite{monoT5}, namely~\OurMethodFirstStage{} and~\OurMethodRerank{}, to show the effectiveness in both dense retrieval and re-ranking. During evaluation, we tested two multi-stage ranking pipelines: (1)~\OurMethodRerank{}: a fine-tuned MonoT5-3B re-ranking BM25 top-100 and (2)~\OurMethodFirstStage{} +~\OurMethodRerank{}: a fine-tuned MonoT5-3B re-ranking a fine-tuned ColBERT top-200 documents.

\subsection{Baselines}

We chose strong competitive rankers as baselines to highlight the effectiveness of our proposed domain adapted ranker.

\textbf{BM25}: Traditional lexical sparse retrieval. We replicated the BM25 scores from scratch.

\textbf{Zero-shot (ZS) Models}: A fine-tuned ranker on MS-MARCO dataset, includes MonoT5-3B and ColBERT.

\textbf{InPars}~\cite{inpars}: An unsupervised training data generation framework for ranking. Synthetic queries are generated from randomly selected documents using few-shot prompting GPT-3 Curie model. Language model likelihood is used as a filtering step to pick top-10k high-quality synthetic queries before fine-tuning any ranker. Based on the reasons provided by~\citet{docgen}, we do not compare against InPars-v2.

\textbf{DocGen-RL}~\cite{docgen}: An RL-driven framework to generate documents from queries. Also an iterative approach, based on expand, highlight, and generate stages, generates documents from queries to prepare training data.

\textbf{Promptagator++}~\cite{promptagator}: As the SOTA methods closest to our work, we evaluate against Promptagator++. This methods operates by randomly selecting 1 million documents from the target collection. It utilizes 8-shot prompting with a 137 billion-parameter FLAN model~\cite{FLAN} to create 8 queries per document. Following consistency filtering, 1 million queries are selected to train a GTR-Base dual-encoder and cross-encoder~\cite{GTR}.

We directly utilized the scores reported by authors for DocGen-RL and Promptagator++. For the remaining baselines, we employed their corresponding HuggingFace~\cite{wolf2020transformers} models to re-run the inference.

\subsection{Tools and Implementation}

Various tools were employed for distinct stages in our pipeline, utilizing Contriever~\cite{contriever} for text encoding, Faiss~\cite{FAISS} for $k$-Means clustering, and Llama2-7B-Chat~\cite{llama2-chat} for query generation, Pyserini for BM25 baseline and hard negative mining, and PyTorch for standard fine-tuning. Throughout our experiments, documents were represented using their title along with the text. Initially, collection documents were filtered for noise by excluding those with a character length less than 300 (can vary across datasets). Greedy decoding with a temperature of 0.0 was employed for the LLM to generate queries.

\subsection{Hyper-Parameter Tuning}

In our methodology section, we introduced several hyper-parameters, all of which underwent tuning to determine the optimal values. These include the temperature $T=1$ (Equation \ref{eq:prob_sampling}), MMR weight $\lambda=1.0$, number of clusters $K=1000$, and training sample size $N=1000$ for ColBERT and $N=1000$ and $5000$ for MonoT5-3B fine-tuning.
We tuned the varying number of in-context examples and found the optimal performance with 3-shot prompting (also used in InPars).
Additionally, through tuning different prompt templates, we discovered that a simple InPars-style template, displayed in Figure \ref{fig:scidocs_prompt}, consistently yields superior retrieval performance across datasets. For the process of hard negative mining, we set the first stage retriever hits $x=100$ and the number of negatives per positive pair $num_{neg}=4$.

We fine-tuned MonoT5-3B using a batch size of 8, gradient accumulation steps of 16, learning rate of $2e^{-5}$, AdamW optimizer with weight decay of 0.01 and warm-up ratio of 0.1, and epochs of 1. To fine-tune ColBERT, we adapted its official pre-training hyper-parameters, including a batch size of 32, a learning rate of $3e^{-6}$, and a maximum sequence length of 300.

The scale and quality of synthetic data depend on the training examples, $N$, and number of clusters, $K$, which we optimize in the subsequent subsections.

\subsubsection{Clustering Optimization}

To represent target domain, we employed $K$-Means algorithm, where $K$ denotes the number of clusters.
We identified the optimal $K$ for each dataset through an unsupervised method, known as the Elbow method~\cite{Thorndike1953WhoBI}. The elbow method computes the Sum of Squared Error (SSE) for each value of $K$, where SSE is calculated as the sum of cosine distances between every collection document and its closest cluster centroid. The optimal $K$ consistently aligns at a fixed point of 1000 across all evaluation datasets, irrespective of variations in corpus size, domain properties, or domain-divergence from MS-MARCO. 

\begin{table*}
    \footnotesize
    \centering
    \resizebox{\textwidth}{!}{
    \begin{tabular}    
    {l|p{5mm}p{3mm}p{4mm}p{5mm}p{6mm}p{4mm}p{5mm}p{4mm}p{5mm}p{4mm}p{5mm}p{4mm}p{4mm}p{5mm}p{6mm}p{5mm}p{6mm}p{7mm}||p{3mm}}

    \toprule        
                
        Datasets (\textrightarrow) & covid & nfc & bio & nq & hotpot & fiqa & signal & news & robust & arg & touché & stack & quora & dbp & scidocs & fever & climate & scifact & avg \\
        
        \midrule
        \midrule[.1em]

        Models ($\downarrow$) & \multicolumn{19}{c}{Retriever} \\
        \midrule

        BM25 & .656 & \textbf{.325} & .465 & .329 & .603 & .236 & \textbf{.330} & .398 & .407 & .315 & \textbf{.367} & .299 & .789 & .313 & \textbf{.158} & .753 & \textbf{.213} & .665 & .423 \\
        Zero-shot ColBERT & .706 & .305 & .480 & .523 & .590 & .318 & .270 & .390 & .392 & .404 & .209 & .350 & .853 & .392 & .144 & .771 & .184 & .672 & .442 \\

        \OurMethodFirstStage{} & \textbf{.751} & \textbf{.325}$^\dagger$ & \textbf{.497}$^\dagger$ & \textbf{.530}$^\dagger$ & \textbf{.614}$^\dagger$ & \textbf{.336}$^\dagger$ & .271 & \textbf{.399} & \textbf{.411}$^\dagger$ & \textbf{.425}$^\dagger$ & .234$^\dagger$ & \textbf{.363}$^\dagger$ & \textbf{.857}$^\dagger$ & \textbf{.401} & .155$^\dagger$ & \textbf{.805}$^\dagger$ & .196$^\dagger$ & \textbf{.688}$^\dagger$ & \textbf{.459} \\
        \midrule

        & \multicolumn{19}{c}{BM25 Top-100 + Re-ranker} \\
        \midrule
        
        Zero-shot MonoT5-3B & .830 & .373 & .559 & .579 & .718 & .462 & .321 & .473 & .566 & .316 & .311 & .421 & .848 & .408 & .193 & .849 & .278 & .760 & .515 \\
        InPars & .803 & - & - & .313 & - & .352 & - & - & .510 & - & - & - & - & .351 & - & - & - & - & - \\
        DocGen-RL & - & - & - & .517 & .663 & - & - & - & - & - & - & - & - & - & - & .720 & - & - & - \\

        \OurMethodRerank{}\textbf{(1k)} & \textbf{.862}$^{\dagger*}$ & \textbf{.382}$^\dagger$ & .588$^\dagger$ & \textbf{.593}$^{\dagger*}$ & \textbf{.748}$^{\dagger*}$ & .458$^*$ & \textbf{.333} & \textbf{.483} & \textbf{.591}$^{\dagger*}$ & .393$^\dagger$ & \textbf{.320} & \textbf{.439}$^\dagger$ & \textbf{.895}$^\dagger$ & \textbf{.422}$^{\dagger*}$ & .200$^\dagger$ & \textbf{.890}$^{\dagger*}$ & \textbf{.310}$^\dagger$ & .757 & \textbf{.537} \\
        
        \OurMethodRerank{}\textbf{(5k)} & .836$^{\dagger*}$ & .376 & \textbf{.590}$^\dagger$ & .588$^{\dagger*}$ & .740$^{\dagger*}$ & \textbf{.465}$^*$ & .300 & .449 & .571$^*$ & \textbf{.427}$^\dagger$ & .269 & \textbf{.439}$^\dagger$ & .894$^\dagger$ & .421$^{\dagger*}$ & \textbf{.202}$^\dagger$ & \textbf{.891}$^{\dagger*}$ & .288$^\dagger$ & \textbf{.761} & .528 \\
        \midrule

        & \multicolumn{19}{c}{Dense Retriever Top-200 + Re-ranker} \\
        \midrule

        GTR (base) retriever & .539 & .308 & .271 & .495 & .535 & .349 & .261 & .337 & .437 & .511 & .205 & .357 & .881 & .347 & .149 & .660 & .241 & .600 & .416 \\

        Promptagator++ & .762 & .370 & - & - & .736 & .494 & - & - & - & \textbf{.630} & \textbf{.381} & - & - & .434 & .201 & .866 & .203 & .731 & .528$^\$$ \\

        ~\OurMethodFirstStage{} & .751 & .325 & .497 & .530 & .613 & .336 & .270 & .399 & .411 & .425 & .234 & .363 & .857 & .403 & .154 & .805 & .196 & .688 & .459 \\

        +~\OurMethodRerank{}\textbf{(1k)} & \textbf{.851} & \textbf{.402} & .594 & \textbf{.671} & \textbf{.769} & .511 & \textbf{.275} & \textbf{.477} & \textbf{.636} & .511 & .331 & .462 & \textbf{.898} & \textbf{.484} & \textbf{.203} & .901 & \textbf{.309} & .758 & \textbf{.558} \\
        +~\OurMethodRerank{}\textbf{(5k)} & .817 & .398 & \textbf{.602} & .661 & .768 & \textbf{.517} & .253 & .422 & .609 & .575 & .262 & \textbf{.463} & .896 & .482 & .202 & \textbf{.903} & .284 & \textbf{.762} & .549 \\

    \bottomrule
    \end{tabular}}
    \caption{Main results comparing nDCG@10 scores between~\OurMethod{} and baselines on BEIR datasets. The best scores across each ranking setting are highlighted in bold. Avg score marked by $^\$$ calculated across only 11 datasets.~\OurMethodRerank{}\textbf{(1k)} and \textbf{(5k)} represent the MonoT5-3B fine-tuned with 1k and 5k training examples correspondingly. Statistical significance reported using two-tailed paired t-test with Bonferroni correction (p < 0.05), against Zero-shot counterparts ($\dagger$) and best of InPars or DocGen-RL ($*$). Promptagator++ was fine-tuned on GTR base, thus we reported GTR scores for comparison.}
    \label{tab:main-ndcg}
\end{table*}

\subsubsection{Optimal Training Sample Size Discovery}

By fixing the optimum number of clusters $K$ at 1000, we determined an optimal training sample size $N$, that proved effective across all datasets. To tune for $N$, we utilized FiQA and NQ as dev datasets, referencing prior work (InPars-v2) which demonstrated improved performance on FiQA and a declined performance on NQ compared to the zero-shot scores. Table \ref{table:optimum_trainsize} displays nDCG@10 values for various instances of $N$, with $K$ fixed at 1000.
Our analysis led us to select optimum $N=1000$ for ColBERT and both $N=1000$ and $5000$ for MonoT5-3B fine-tuning across the datasets.

\begin{table}
\begin{center}
    \resizebox{.35\textwidth}{!}{
    \begin{tabular}{ c r|l l }
        \toprule
        & $N$ & FiQA & NQ \\
        \hline \hline
        \multirow{4}{7em}{MonoT5-3B} & (ZS) 0 & .4617$\dagger$ & .5792 \\
        \cmidrule{3-4}
        & 1k & .4581 & \textbf{.5934} \\ 
        & 5k & \textbf{.4646} & .5880$\dagger$ \\ 
        & 10k & .4553 & .5777 \\
        \cmidrule{1-4}
        \multirow{4}{7em}{ColBERT} & (ZS) 0 & .3183 & .5228 \\
        \cmidrule{3-4}
        & 1k & .3356$\dagger$ & \textbf{.5301} \\ 
        & 5k & \textbf{.3388} & .5233$\dagger$ \\ 
        & 10k & .3306 & .5171 \\
        \bottomrule
    \end{tabular}
    }
    \caption{Ranking performances evaluated on nDCG@10 across the scale of training sample size $N$ on dev datasets. Bold and $\dagger$ indicate the best and second-best scores across benchmarks for each ranker.}
    \label{table:optimum_trainsize}
\end{center}
\vspace{-13pt}
\end{table}

\section{Results and Discussion}
\label{results}

In this section, we present our main experimental results and delve into the key observations.
We first describe our primary findings, reported using nDCG@10 in Table \ref{tab:main-ndcg} comparing between baselines and our approach within each ranking setting.
Second, we report the first-stage retrieval performances, measured using R@100 in Table \ref{tab:main-recall}.

\subsection{Re-ranking Results}

\begin{table*}
\footnotesize
\begin{center}
    \resizebox{\textwidth}{!}{%
    \begin{tabular}{l|p{3mm}p{3mm}p{3mm}p{3mm}p{3mm}p{3mm}p{3mm}p{3mm}p{3mm}p{3mm}p{3mm}p{3mm}p{3mm}p{3mm}p{3mm}p{3mm}p{3mm}p{5mm}||l}
        \toprule

        \shortstack{Datasets (\textrightarrow)\\ Models ($\downarrow$)} & \rotatebox{90}{covid}  & \rotatebox{90}{nfc}  & \rotatebox{90}{bio}  & \rotatebox{90}{nq}  & \rotatebox{90}{hotpot}  & \rotatebox{90}{fiqa}  & \rotatebox{90}{signal}  & \rotatebox{90}{news}  & \rotatebox{90}{robust}  & \rotatebox{90}{arg}  & \rotatebox{90}{touché}  & \rotatebox{90}{stack}  & \rotatebox{90}{quora}  & \rotatebox{90}{dbp}  & \rotatebox{90}{scidocs}  & \rotatebox{90}{fever}  & \rotatebox{90}{climate}  & \rotatebox{90}{scifact}  & \rotatebox{90}{\textbf{avg}} \\

        \midrule
        \midrule[.1em]
        
        BM25 & .498 & .250 & \textbf{.714} & .760 & .740 & .540 & \textbf{.370} & \textbf{.422} & \textbf{.375} & \textbf{.942} & \textbf{.538} & .606 & .973 & .398 & .356 & .931 & .436 & \textbf{.908} & .598 \\
        
        Zero-shot ColBERT & .473 & .255 & .664 & .911 & .747 & .598 & .278 & .369 & .311 & .885 & .436 & .625 & .989 & .458 & .345 & .934 & .447 & .878 & .589 \\

        \midrule

        \textbf{\OurMethodFirstStage{}} & \textbf{.544}$^\dagger$ & \textbf{.272}$^\dagger$ & .691$^\dagger$ & \textbf{.915} & \textbf{.769}$^\dagger$ & \textbf{.615}$^\dagger$ & .291 & .380$^\dagger$ & .321$^\dagger$ & .906$^\dagger$ & .474$^\dagger$ & \textbf{.645}$^\dagger$ & \textbf{.990} & \textbf{.493}$^\dagger$ & \textbf{.356}$^\dagger$ & \textbf{.948}$^\dagger$ & \textbf{.465}$^\dagger$ & .899$^\dagger$ & \textbf{.610} \\

        \bottomrule
    \end{tabular}}
    \caption{Comparison of R@100 scores across baselines and \OurMethod{}. The best scores for each dataset are highlighted in bold. Statistical significance reported using two-tailed paired t-test with Bonferroni correction (p < 0.05), against Zero-shot counterpart ($\dagger$).}
    \label{tab:main-recall}
\end{center}
\end{table*}

In Table \ref{tab:main-ndcg}, it is evident that~\OurMethod{} consistently surpasses the SOTA baselines in most cases, exhibiting notable improvements in performance. Specifically,~\OurMethod{} consistently and substantially outperforms both InPars and DocGen-RL re-rankers, showcasing average relative enhancements of 26\% and 17\% respectively across the evaluation datasets they share. When compared to Promptagator++,~\OurMethod{} demonstrates an average relative improvement of 4\% across the shared evaluation datasets. Remarkably,~\OurMethod{} surpasses Promptagator++ in performance, utilizing merely 1000 LLM calls and fine-tuning with only 1000 training pairs, in contrast to Promptagator++'s requirement of generating 8 million queries using a 137B LLM and fine-tuning with 1 million training pairs. This highlights the effectiveness of our efficient and robust approach compared to the complex, resource-intensive, and exhaustive training methods based on reinforcement learning.

In many instances, the performance of the SOTA baselines degraded, compared to zero-shot counterparts. For instance, both InPars and DocGen-RL consistently demonstrate performance decreases relative to the zero-shot MonoT5-3B, with Avg. decrements of 18\% and 11\% respectively across the evaluation datasets they share (DocGen-RL also underperforms compared to zero-shot MonoT5-base, as shown in Table ~\ref{tab:models-ablations}).
On the other hand,~\OurMethod{} consistently surpasses all zero-shot models across all BEIR datasets, whether trained with 1,000 or 5,000 synthetic training examples.

Interestingly, training~\OurMethodRerank{} with only 1,000 synthetic examples exhibited a slight performance improvement compared to training with 5,000 synthetic examples on 13 out of 18 datasets, indicating the sample efficiency of our approach. In the future, it may be feasible to automatically determine the minimum training size ($N$) for each dataset or task.

\subsection{First-Stage Retrieval Results}

In Table \ref{tab:main-recall}, similar to nDCG@10 scores, R@100 also demonstrates more substantial improvements for larger domain-shifts (7.1\%\footnote{denotes absolute precentage improvement.\label{abs_imp}} on TREC-COVID and 3.8\%$^{\ref{abs_imp}}$ on Touché-2020) and limited improvements for smaller domain-shifts (.4\%$^{\ref{abs_imp}}$ on NQ). On average,~\OurMethod{} enhances zero-shot ColBERT by 2.1\%$^{\ref{abs_imp}}$ on BEIR datasets.

\begin{table}
\vspace{-0.3cm}
\begin{center}
    \resizebox{0.48\textwidth}{!}{%
    \begin{tabular}{ r c r|l l }
        \toprule
        LLM & Prompt & Size & FiQA & NQ \\
        \midrule\midrule[.1em]

        (Zero-shot) No LLM & & - & .3702 & .5404 \\
        \cmidrule{1-5}
        \multirow{2}{*}{LLAMA-2 7B Chat} &
        ms-marco & 7B & .3736 & .5371 \\
        & in-domain & 7B & .3811 & .5444 \\
        \cmidrule{1-3}
        LLAMA-2 13B Chat & 
        \multirow{4}{*}{in-domain}
        & 13B & \textbf{.3912} & .5370 \\
        BLOOM-3B & & 3B & .3380 & .5193 \\
        BLOOM-7B1 & & 7.1B & .3634 & .5172 \\
        gpt-3.5-turbo & & 20B & .3742 & \textbf{.5466} \\
        
        \bottomrule
    \end{tabular}}
    \caption{nDCG@10 performances across different LLMs for query generation ($K=1$, $N=5000$) with MonoELECTRA re-ranker fine-tuned on the generations}
    \label{table:query_generators}
\end{center}
\vspace{-0.5cm}
\end{table}

\begin{table*}[htpb]
\resizebox{0.5\textwidth}{!}{%
\begin{tabular}{cc|c|c|c|c|c|c|}
\cline{3-8}

& \multirow{2}{*}{10k} & .391 & .391 & .399 & .396 & .398 & .389 \\
& & \relup{2} & \relup{2} & \relup{3} & \relup{3} & \relup{3} & \relup{2} \\
\cline{3-8}

& \multirow{2}{*}{5k} & .381 & .390 & .395 & \textbf{.392} & .387 & \multirow{2}{*}{-} \\
& & \relup{1} & \relup{2} & \relup{3} & \textbf{\relup{2}} & \relup{2} & \\
\cline{3-8}

& \multirow{2}{*}{1k} & .382 & .385 & .390 & .384 & \multirow{2}{*}{-} & \multirow{2}{*}{-} \\
& & \relup{1} & \relup{2} & \relup{2} & \relup{1} & & \\
\cline{3-8}

& \multirow{2}{*}{500} & .381 & .381 & .381 & \multirow{2}{*}{-} & \multirow{2}{*}{-} & \multirow{2}{*}{-} \\
& & \relup{1} & \relup{1} & \relup{1} & & & \\
\cline{3-8}

\multicolumn{1}{c}{\multirow{-5}{*}{\begin{tabular}[c]{@{}c@{}}Number of \\ Training \\ Examples \\ (N)\end{tabular}}} & \multirow{2}{*}{100} & .359 & .371 & \multirow{2}{*}{-} & \multirow{2}{*}{-} & \multirow{2}{*}{-} & \multirow{2}{*}{-} \\
& & \reldown{1} & \relup{0} & & & & \\
\cline{3-8}

 & \multicolumn{1}{c}{} & \multicolumn{1}{c}{1} & \multicolumn{1}{c}{100} & \multicolumn{1}{c}{500} & \multicolumn{1}{c}{1k} & \multicolumn{1}{c}{5k} & \multicolumn{1}{c}{10k} \\ 
 & \multicolumn{1}{c}{} & \multicolumn{6}{c}{\shortstack{Number of Clusters (K)\\ \\ Zero-shot MonoELECTRA score on FiQA is 0.370}} \\ 
\end{tabular}}
\resizebox{0.5\textwidth}{!}{%
\begin{tabular}{cc|c|c|c|c|c|c|}
\cline{3-8}

& \multirow{2}{*}{10k} & .532 & .547 & .542 & .546 & .540 & .546 \\
& & \reldown{1} & \relup{1} & \relup{0} & \relup{1} & \relup{0} & \relup{1} \\
\cline{3-8}

& \multirow{2}{*}{5k} & .544 & .542 & .544 & \textbf{.551} & .543 & \multirow{2}{*}{-} \\
& & \relup{0} & \relup{0} & \relup{0} & \textbf{\relup{1}} & \relup{0} &  \\
\cline{3-8}

& \multirow{2}{*}{1k} & .546 & .547 & .550 & .548 & \multirow{2}{*}{-} & \multirow{2}{*}{-} \\
& & \relup{1} & \relup{1} & \relup{1} & \relup{1} &  &  \\
\cline{3-8}

& \multirow{2}{*}{500} & .537 & .548 & .548 & \multirow{2}{*}{-} & \multirow{2}{*}{-} & \multirow{2}{*}{-} \\
& & \reldown{0} & \relup{1} & \relup{1} &  &  &  \\
\cline{3-8}

\multicolumn{1}{c}{\multirow{-5}{*}{\begin{tabular}[c]{@{}c@{}}Number of \\ Training \\ Examples \\ (N)\end{tabular}}} & \multirow{2}{*}{100} & .539 & .546 & \multirow{2}{*}{-} & \multirow{2}{*}{-} & \multirow{2}{*}{-} & \multirow{2}{*}{-} \\
& & \reldown{0} & \relup{1} &  &  &  &  \\
\cline{3-8}

 & \multicolumn{1}{c}{} & \multicolumn{1}{c}{1} & \multicolumn{1}{c}{100} & \multicolumn{1}{c}{500} & \multicolumn{1}{c}{1k} & \multicolumn{1}{c}{5k} & \multicolumn{1}{c}{10k} \\ 
 & \multicolumn{1}{c}{} & \multicolumn{6}{c}{\shortstack{Number of Clusters (K)\\ \\ Zero-shot MonoELECTRA score on NQ is 0.540}} \\ 
\end{tabular}}
\caption{Fine-tuned MonoELECTRA re-ranking performances in nDCG@10 for different values of $K$ and $N$ on FiQA (left) and NQ (right). \relup{} and \reldown{} indicate integer rounded superior or inferior performance percentage against zero-shot scores.}
\label{tab:fiqa-table}
\vspace{-13pt}
\end{table*}

\section{Analysis}
\label{analysis}

In this section, we report our analysis of~\OurMethod{}'s performance, which includes examining the need for clustering, confirming the choice of the query generator, and validating the quality of the generated queries.

\subsection{Effect of Clustering for Domain Adaptation}
\label{sec:effect_clustering}

We employ clustering to represent the target domain and number of training samples to force diversity during fine-tuning. However, we question whether clustering genuinely contributes to the process and, if so, how it influences the overall performance. Additionally, we take the training sample size $N$, into account. In Table~\ref{tab:fiqa-table}, we illustrate the combined effect of both the $K$ and $N$ on MonoELECTRA top-100 BM25 re-ranking performances, measured in nDCG@10. MonoELECTRA is used in the analysis Sections \ref{sec:effect_clustering} and \ref{sec:effect_querygen} in order to measure the amplified performance improvements in a smaller model, as described in the previous section.

Table~\ref{tab:fiqa-table} confirms our decision to select $N=5000$ for MonoELECTRA.
Notably, this figure highlights that the most substantial and consistent improvements occur around the values of \{K=1000, N=5000\} across both datasets. Performances without clustering ($K=1$) often fall below zero-shot in both datasets, especially NQ exhibiting the poorest performances.

\subsection{Effect of Query Generators}
\label{sec:effect_querygen}

We conducted an ablation study on query generation to assess how the quality of generated queries impacts overall retrieval performance. Table \ref{table:query_generators} displays the performances of MonoELECTRA fine-tuned with queries generated by various LLMs, including LLAMA2-Chat (7B and 13B), BLOOM (3B and 7B), and GPT-3.5-turbo~\cite{GPT}.

In comparison to the zero-shot re-ranking scores, LLAMA-2 7B was deemed the optimal choice for our query generator.
LLAMA-2 7B with 3-shot in-domain prompts exhibits higher improvements on both dev datasets, surpassing gpt-3.5-turbo.
While LLAMA-2 13B demonstrates superior performance to 7B on FiQA, it falls below the zero-shot performance in NQ, attributed to its large model capacity and sensitivity to prompts~\cite{calibrate-LLM}.
BLOOM generates short queries lacking context, despite having sufficient contextual query examples from 3-shot examples. 
GPT-3.5-turbo generates high-quality queries, resulting in improved performance over zero-shot, but tends to be unstable with few-shot prompts, suggesting potential for further prompt engineering to enhance performance on each dataset. 
Our second main contribution involves using in-domain 3-shot prompts to generate queries over the ms-marco prompt, showcasing notable improvements on LLAMA-2 7B model. 

\subsection{Examples of~\OurMethod{} Queries}
\label{sec:example_queries}

\begin{figure}[ht]
\vspace{-0.5cm}
    \begin{center}
        \begin{subfigure}{0.5\textwidth}
             \centering
             \includegraphics[width=\textwidth]{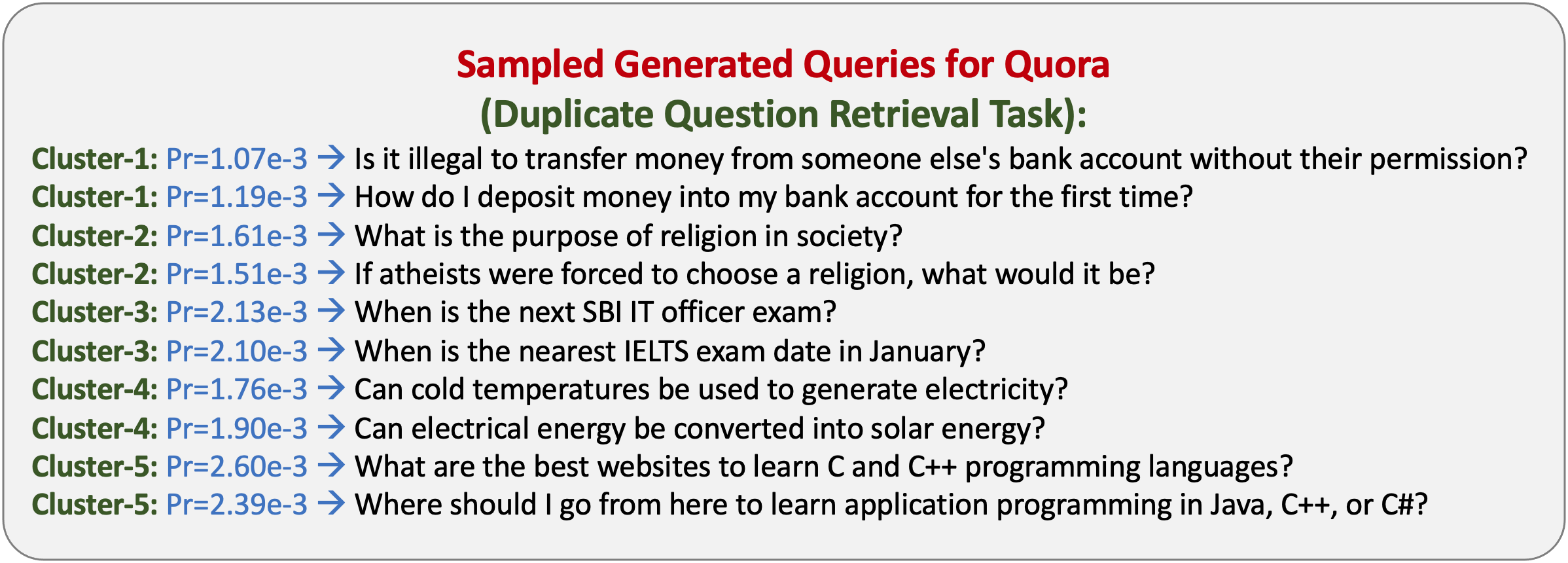}
             \caption{}
             \label{fig:queries_quora}
         \end{subfigure}
     
         \begin{subfigure}{0.5\textwidth}
             \centering
             \includegraphics[width=\textwidth]{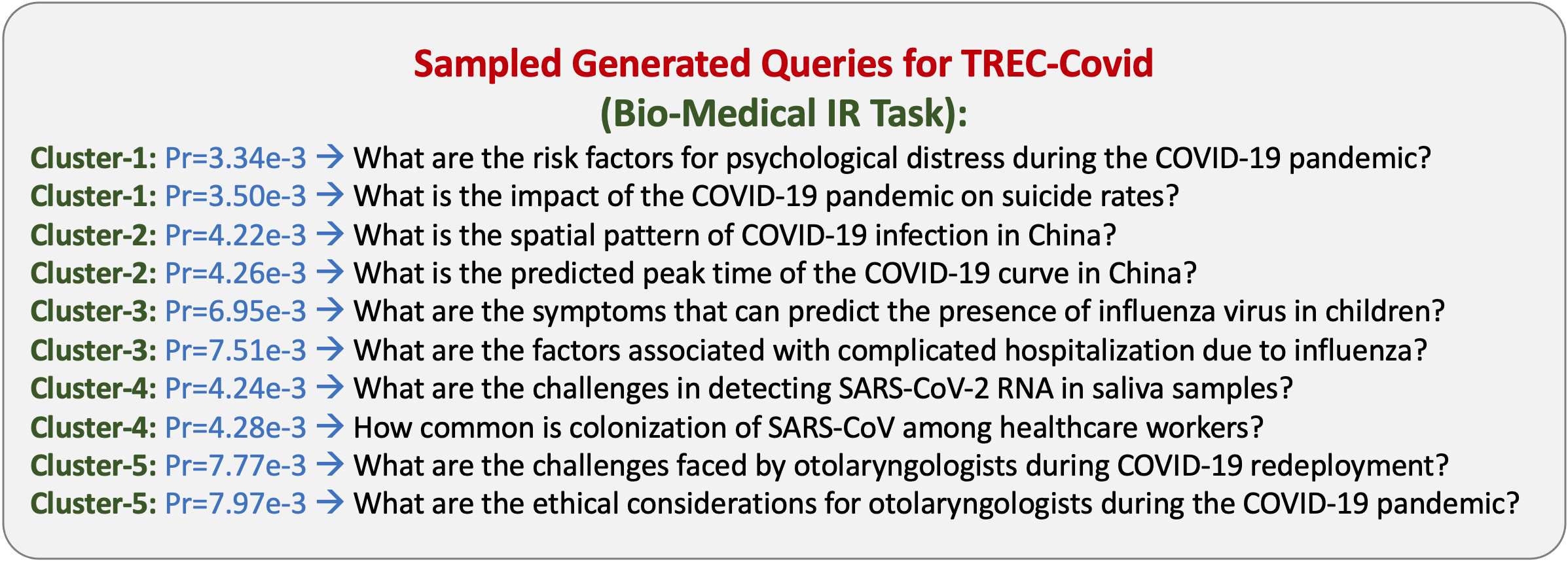}
             \caption{}
             \label{fig:queries_covid}
         \end{subfigure}
     
    \caption{Example queries generated by~\OurMethod{} on (a) Quora and (b) TREC-Covid datasets. Pr denotes the $Pr(D_i|cluster_k)$ where $D_i$ and $cluster_k$ refer to $i^{th}$ document and $k^{th}$ cluster.}
    \label{fig:example_queries}
    \end{center}
    \vspace{-0.5cm}
\end{figure}

So far, we have evaluated the effectiveness of~\OurMethod{} using quantitative measures and are now shifting our focus to examining the actual queries produced by our method. Figure~\ref{fig:example_queries} presents ten example queries generated from the Quora and TREC-Covid datasets, each representing distinct tasks and domains. In Figure~\ref{fig:example_queries}, the synthetic queries are sampled across different clusters with different probability scores $Pr(D_i|cluster_k)$. For instance, in Figure~\ref{fig:queries_quora}, we observe that in the Quora duplicate question retrieval task, each cluster corresponds to sub-topics of the target domain representation, such as monetary bank transfers, religion, exams in India, energy, and programming languages. Within each cluster, diverse queries are sampled using different probabilistic scores to aid in learning the domain representation. Additionally, the generated queries contain sufficient context or entities to retrieve pertinent information from its respective collection. This analysis of the generated queries further validates the effectiveness of our approach in generating a diverse and representative set of high-quality queries.

\section{Conclusions}
\label{conclusion}

We proposed a general unsupervised domain adaptation method ~\OurMethod, which can be used to fine-tune any ranking model for a given target domains. \OurMethod{} introduced significant innovations over the previously reported unsupervised domain adaptation methods. Specifically, \OurMethod{} proposes representing the target domain collection with document clustering; an effective method to diversify the synthetically generated queries, and an effective prompting strategy for using an LLM to generate more effective and representative synthetic training data. We experimentally demonstrated that~\OurMethod{} is both scalable and effective, as it uses only a few thousands of synthetic training examples, while consistently improves over the SOTA zero-shot rankers, and significantly outperforms the SOTA methods for unsupervised domain adaptation methods in most cases. We complemented the strong empirical performance of~\OurMethod{} with an in-depth analysis of the components to quantify their contributions. Together, the presented techniques and experimental results significantly advance neural ranking adaptation, establish a new state-of-the-art in neural ranking, and suggest promising directions for future improvements. 

\section{Limitations}

Our proposed methodology involves two pivotal steps: (1) clustering; and (2) query generation. First, we employ Contriever as our text encoder to produce embeddings for clustering. While we anticipate that it will produce high-quality document representation and prove to be useful in our work, we did not assess other document embeddings. Future work could directly address the question of choosing the appropriate embedding for clustering.

Secondly, we employed the Faiss library to implement $K$-Means clustering. However, as the collection size scales up over the millions, clustering becomes impractical. Consequently, Faiss resorts to sampling the collection and then training their algorithm. This loss of information during sampling could propagate as errors in the final retrieval scores. However, given that large collections typically contain dense clusters, the process of sampling for clustering in such cases may pose less problem.

Akin to many previous studies~\cite{calibrate-LLM}, we often encountered a lack of robustness of LLMs and their sensitivity to minor changes in the prompt affecting subsequent retrieval performance. Future work could explore strategies to mitigate this robustness through techniques like calibration~\cite{calibrate-LLM} and perform corresponding studies to see the impact on reranking.

\section{Ethical Considerations}
Retrieval systems may give rise to a variety of ethical issues, such as the potential for bias, which can result in the preferential treatment of specific perspectives, a lack of transparency due to the opaque nature of deep learning models, obscuring the reasons behind the ranking of documents, and, in extreme cases, the facilitation of echo chambers. Therefore, it is essential to conduct thorough testing of these systems both prior to and during their deployment.

As shown by our work, the performance of downstream retrievers can be rightly influenced by the LLMs employed for generating synthetic queries. Given that LLMs can produce content that is inaccurate or entirely fabricated, there's a risk that they might generate problematic queries, especially if applied to sensitive datasets. Although this issue may appear less critical in a scenario such as ours where the generated content is intended solely as training material for a following retriever, there is still a potential for generating harmful and toxic queries. Such queries could lead the retriever towards biased outcomes. Therefore, it is imperative to assess these systems to mitigate these risks against biases of the data generator.

\section*{Acknowledgements}
The authors are thankful to Harshita Sahijwani and Sergey Volokhin, and the reviewers and meta-reviewers for their valuable comments and suggestions. This work was supported in part by the IARPA BETTER program (\#2019-19051600005). The views and conclusions contained in this work are those of the authors and should not be interpreted as necessarily representing the official policies, either expressed or implied, or endorsements of ODNI, IARPA, or the
U.S. Government. The U.S. Government is authorized to reproduce and distribute reprints for governmental purposes notwithstanding any copyright annotation therein.

\bibliography{acl_latex}
\appendix
\section{Appendix}
\label{sec:appendix}

\begin{table*}
\footnotesize
    \centering
    \resizebox{\textwidth}{!}{
    \begin{tabular}{l | c|cc|cc|cc|cc}
    \toprule

        \multicolumn{2}{c|}{} & \multicolumn{2}{c|}{Dense Retriever} & \multicolumn{6}{c}{Re-ranker using BM25 Top-100} \\

        \cmidrule(rl){3-4} \cmidrule(rl){5-10}

        \multicolumn{1}{c}{Size (\textrightarrow)} & \multicolumn{1}{c|}{} & \multicolumn{2}{c|}{110M} & \multicolumn{2}{c|}{110M} & \multicolumn{2}{c|}{220M} & \multicolumn{2}{c}{3B}  \\
        \cmidrule(rl){2-2} 
        \cmidrule(rl){3-4} \cmidrule(rl){5-10}

        Ranker (\textrightarrow) & \multirow{3}{*}{BM25} & \multicolumn{2}{c|}{ColBERT} & \multicolumn{2}{c|}{MonoELECTRA} & \multicolumn{2}{c|}{MonoT5-base} & \multicolumn{2}{c}{MonoT5-3B} \\

        \cmidrule(rl){3-4} \cmidrule(rl){5-6} \cmidrule(rl){7-8} \cmidrule(rl){9-10}

         \multirow{2}{*}{Datasets ($\downarrow$)} &  & \multirow{2}{*}{Zero-shot} & \multirow{2}{*}{\textbf{\shortstack{DUQGen-\\retriever}}} & \multirow{2}{*}{Zero-shot} & \multirow{2}{*}{\textbf{\shortstack{DUQGen-\\reranker(5k)}}} & \multirow{2}{*}{Zero-shot} & \multirow{2}{*}{\textbf{\shortstack{DUQGen-\\reranker(5k)}}} & \multirow{2}{*}{Zero-shot} & \multirow{2}{*}{\textbf{\shortstack{DUQGen-\\reranker(5k)}}} \\

        & & & & & & & & & \\

        \midrule
        \midrule[.1em]

        covid  & .656 & .706 & .751\relup{6} & .730 & .761\relup{4} & .814 & \textbf{.853}\relup{5} & .830 & .836\relup{1} \\
        nfc  & .325 & .305 & .325\relup{6} & .280 & .356\relup{27} & .357 & .368\relup{3} & .373 & \textbf{.376}\relup{1} \\
        bio  & .465 & .480 & .497\relup{4} & .502 & .523\relup{4} & .531 & .566\relup{7} & .559 & \textbf{.590}\relup{6} \\
        nq  & .329 & .523 & .530\relup{1} & .540 & .551\relup{2} & .540 & .550\relup{2} & .579 & \textbf{.588}\relup{2} \\
        hotpot  & .603 & .590 & .614\relup{4} & .691 & .709\relup{3} & .698 & .721\relup{3} & .718 & \textbf{.740}\relup{3} \\
        fiqa  & .236 & .318 & .336\relup{5} & .370 & .392\relup{6} & .391 & .400\relup{2} & .462 & \textbf{.465}\relup{1} \\
        signal  & .330 & .270 & \textbf{.271}\relup{0} & .297 & .284\reldown{-4} & .316 & .309\reldown{-2} & .321 & .300\reldown{-6} \\
        news  & .398 & .390 & .399\relup{2} & .445 & .411\reldown{-8} & .459 & \textbf{.470}\relup{2} & .473 & .449\reldown{5} \\
        robust  & .407 & .392 & .411\relup{5} & .440 & .479\relup{9} & .518 & .538\relup{4} & .566 & \textbf{.571}\relup{1} \\
        arg  & .315 & .404 & .425\relup{5} & .233 & .327\relup{40} & .188 & .383\relup{103} & .316 & \textbf{.427}\relup{35} \\
        touché  & .367 & .209 & .234\relup{12} & .278 & .261\reldown{-6} & .305 & \textbf{.347}\relup{14} & .311 & .269\reldown{-13} \\
        stack  & .299 & .350 & .363\relup{4} & .339 & .387\relup{14} & .389 & .405\relup{4} & .421 & \textbf{.439}\relup{4} \\
        quora  & .789 & .853 & .857\relup{0} & .730 & .873\relup{20} & .845 & .888\relup{5} & .848 & \textbf{.894}\relup{5} \\
        dbp  & .313 & .392 & .401\relup{2} & .278 & .389\relup{40} & .395 & .406\relup{3} & .408 & .\textbf{421}\relup{3} \\
        scidocs  & .158 & .146 & .155\relup{6} & .162 & .182\relup{12} & .171 & .186\relup{9} & .193 & \textbf{.202}\relup{5} \\
        fever  & .753 & .771 & .805\relup{4} & .816 & .867\relup{6} & .826 & .878\relup{6} & .849 & \textbf{.891}\relup{5} \\
        climate  & .213 & .184 & .196\relup{7} & .246 & \textbf{.296}\relup{21} & .251 & .268\relup{7} & .278 & .288\relup{3} \\
        scifact  & .665 & .672 & .688\relup{2} & .684 & .727\relup{6} & .730 & .746\relup{2} & .760 & \textbf{.761}\relup{0} \\
        \midrule
        \midrule
        \textbf{avg} & .423 & .442 & .459\relup{4} & .448 & .487\relup{9} & .485 & .516\relup{6} & .515 & \textbf{.528}\relup{3} \\

    \bottomrule
    \end{tabular}}
    \caption{Comparison of nDCG@10 scores across different model sizes and different ranking families. The best scores are highlighted in bold. \textcolor{blue}{Blue} and \textcolor{red}{red} colored percentage values indicate the relative improvements in performance compared to the corresponding zero-shot baseline. Suffix \textbf{(5k)} refers to the training size used to fine-tune corresponding models.}
    \label{tab:models-ablations}
\end{table*}

\subsection{Effect on Model Sizes}

Different model sizes and different ranker families have been shown to exhibit different performances. Therefore, we fine-tuned MonoELECTRA\footnotemark{}~\cite{electra}\footnotetext{cross-encoder/ms-marco-electra-base}, MonoT5-base\footnotemark{}\footnotetext{castorini/monot5-base-msmarco-10k}, MonoT5-3B, and ColBERT to show the effectiveness of our approach against the model sizes. We pre-trained the MonoELECTRA re-ranker on MS-MARCO using a batch size of 32, learning rate of $2e^{-5}$, AdamW optimizer with weight decay of 0.01 and warm-up ratio of 0.1, regression loss, and a maximum sequence length of 512. We continued using the same hyper-parameters for fine-tuning too. For MonoT5-base fine-tuning we only changed the batch size to 8 and kept the remaining hyper-parameters same as of MonoELECTRA.

The complete performances of~\OurMethod{} against each zero-shot baseline measured in nDCG@10 is reported in Table \ref{tab:models-ablations}. \OurMethod{} shows consistent improvements over corresponding zero-shot baselines across almost all datasets and across all model sizes. The larger the model size, the higher the performance improvements are, for example, the average score across all datasets increases from .487 to .528 between MonoELECTRA and MonoT5-3B. It is important to note that the performance improvements achieved by~\OurMethod{} are larger (4\% Avg.) in smaller models and smaller (1\% Avg.) in larger models. Fine-tuning on ColBERT shows consistent improvements across all datasets with no drop in performance. Thus showcasing the robustness of our approach to deploying dense retrievers in practical systems.

\subsection{Running Time of~\OurMethod{}}

\OurMethod{} is a cost-effective and easily scalable to a large corpus size. ~\OurMethod{} needs to apply the pipeline only once per domain to generate training data to fine-tune a ranker. Then, at inference time, no additional complexity overheard is added. The training dataset generation process is a sequence of three steps, each fully controllable by a small set of parameters, and can be scaled to a large document collection as our experiments show. In fact, the evaluation datasets of NQ, FEVER, BioASQ are 2.5 Million, 5.5 Million, and 14.9 Million documents respectively. To give an idea of the time complexity, single-pass clustering of the full NQ collection required 8 minutes\footnote{run on Quadro RTX 8000 GPU with 48GB memory.\label{machine1}} on a single machine; query generation using LLama2-7B to create $N=5,000$ synthetic training examples was the slowest step which required 1.2 hours\footnote{run on NVIDIA H100 GPU with 80GB memory.\label{machine2}}, and can be increased or reduced easily by modifying N. The fine-tuning of the largest ranker model MonoT5-3B required 38 minutes$^{\ref{machine2}}$.

\subsection{Query Generation Prompts}

\begin{figure*}
\begin{center}

    \begin{subfigure}[b]{.49\textwidth}
         \centering
         \includegraphics[width=\textwidth]{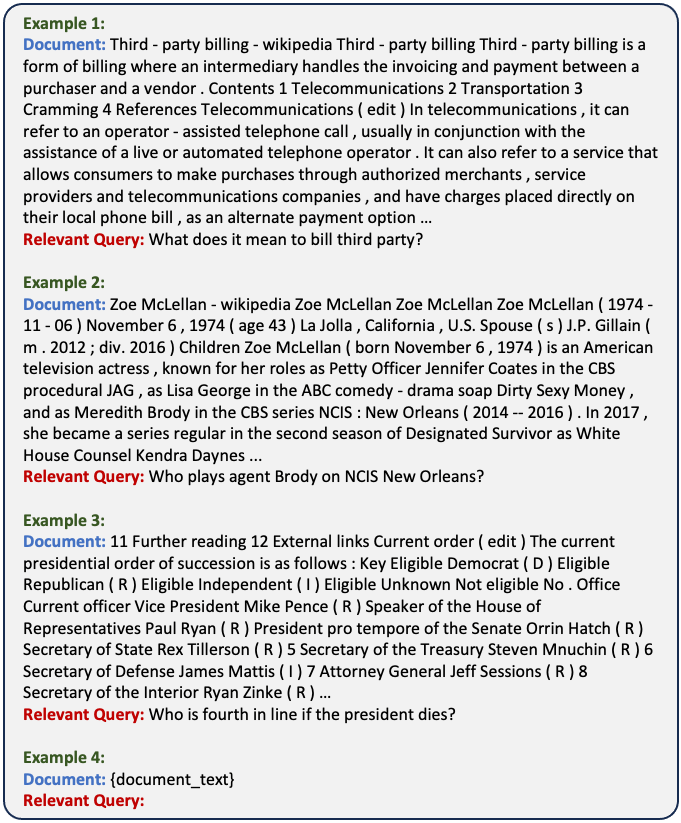}
         \caption{}
         \label{fig:nq_prompt}
     \end{subfigure}
     \begin{subfigure}[b]{.49\textwidth}
         \centering
         \includegraphics[width=\textwidth]{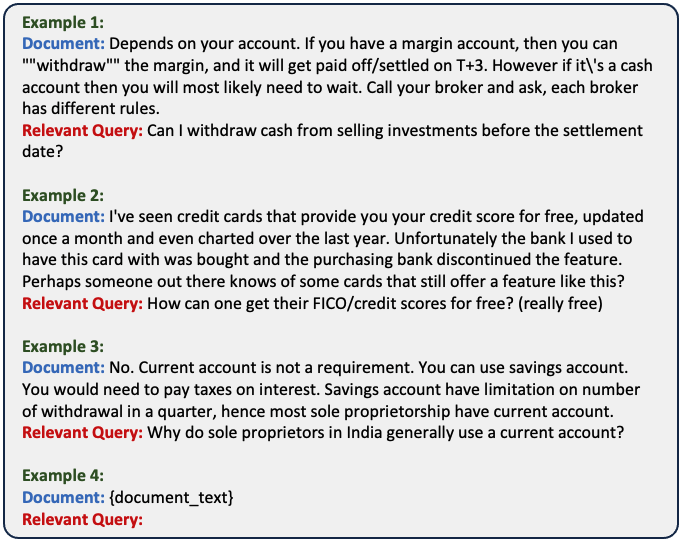}
         \caption{}
         \label{fig:fiqa_prompt}
     \end{subfigure}

    \caption{Example prompts used for (a) NQ and (b) FiQA dataset.}
    \label{fig:nq_and_fiqa_prompt}
    
\end{center}
\end{figure*}

Using additional prompts can be helpful in showcasing the diversity of tasks in BEIR and highlighting the minimal effort required for any domain using our approach.~\OurMethod{} can be utilized across a wide range of retrieval tasks with significantly reduced human effort, requiring only a few labeled examples as demonstrations. Thus, in Figures \ref{fig:nq_prompt} and \ref{fig:fiqa_prompt}, we present the additional prompts utilized for dev datasets (NQ and FiQA) to exemplify the diverse tasks and the minimal effort involved.

\end{document}